\documentclass{aa}  
\usepackage[utf8]{inputenc}
\usepackage{graphicx}
\usepackage{txfonts}
\usepackage{mathtools}
\usepackage{nameref}
\usepackage{wasysym}
\usepackage{makecell}
\usepackage[caption=false]{subfig}
\usepackage{hyperref}
\usepackage{xcolor}
\usepackage{float}
\hypersetup{
    colorlinks=true,
    linkcolor=blue,
    citecolor=blue,
    filecolor=magenta,      
    urlcolor=blue,
}
\usepackage[capitalise]{cleveref}
\usepackage{natbib}
\usepackage{soul}
\bibpunct{(}{)}{;}{a}{}{,} 

\newcommand{\corr}[2]{#2}

\begin{document}

\title{The challenge of forming a fuzzy core in Jupiter}

\author{Simon Müller\inst{1}
        ,
        Ravit Helled\inst{1}
        \and
        Andrew Cumming\inst{2}
       }

\institute{Center for Theoretical Astrophysics and Cosmology \\
           Institute for Computational Science, University of Zürich \\
           Winterthurerstrasse 190, 8057 Zürich, Switzerland \\
          \email{muesim@physik.uzh.ch}
          \and
           Department of Physics and McGill Space Institute, McGill University, \\
           3600 rue University, Montreal, QC H3A 2T8, Canada
          }
\authorrunning{Müller et al.}

\date{Received 20 December 2019 / Accepted 27 April, 2020}

\abstract{
Recent structure models of Jupiter that match Juno gravity data suggest that the planet harbours an
extended region in its deep interior that is enriched with heavy elements: a so-called dilute \corr{fuzzy}{or fuzzy} core.
This finding raises the question of what possible formation pathways could have lead to such a structure.
We modelled Jupiter's formation and long-term evolution, starting at late-stage formation before runaway gas accretion. 
The formation scenarios we considered include both primordial composition gradients, as well as gradients  that are built 
as proto-Jupiter rapidly acquires its gaseous envelope. We then \corr{follow}{followed} Jupiter's evolution as it cools down and contracts,
with a particular focus on the energy and material transport in the interior.
We find that none of the scenarios we consider lead to a fuzzy core that is compatible with
interior structure models. In all the cases, most of Jupiter's envelope becomes convective and fully mixed after a few million years at most. This is true even when we \corr{consider}{considered} a case where the gas accretion leads to a cold planet,
and large amounts of heavy elements are accreted.
We therefore conclude that it is very challenging to explain Jupiter's dilute core from standard formation models. 
We suggest that future works should consider more complex formation pathways as well as the modelling of additional physical 
processes that could lead to Jupiter's current-state internal structure.}


\maketitle

\section{Introduction}\label{sec:introduction}
The Juno mission \citep{Bolton2017} recently mapped Jupiter's gravitational field with high precision \citep{Folkner2017,Iess2018}. 
Traditionally, Jupiter's interior \corr{has been}{was} modelled with a three-layer structure including:  (i) a central compact icy and rocky
core; (ii) an inner envelope of metallic hydrogen and helium; and (iii) an outer envelope of molecular hydrogen and helium. These models 
typically assumed an adiabatic temperature profile for the planet, and the distribution of the heavy elements within the envelope(s) was 
assumed to be constant (e.g. \citet{Stevenson1982,Guillot2005}). New interior structure models that fit the gravitational moments suggest 
that Jupiter has a \corr{diluted/fuzzy core}{dilute or fuzzy core}, rather than a compact one \citep{Wahl2017,Debras2019}. This suggests 
that there is an extended region in the innermost part of the planet that is highly enriched with heavy elements.

The idea that Jupiter could have heavy-element composition gradients was proposed decades ago \citep{Stevenson1982,Stevenson1985},
and there has been an ongoing effort to explore structure models with chemically in-homogenous layers and non-adiabatic interiors 
\citep{Chabrier2007,Leconte2012,Lozovsky2017,Berardo2017a,Vazan2018,Debras2019}.
Recent formation models confirm that giant planets are expected to form with a primordial composition gradient in their deep interiors,
but this region is typically more compact than predicted by structure models \citep{Lozovsky2017,Helled2017}.
An interior with a composition gradient 
could be a result of core erosion (e.g. \citet{Guillot2004, Moll2017}, by which heavy elements from a compact core are dredged
upwards. Constraints on the transport properties of heavy elements in \corr{hydrogen/helium}{hydrogen-helium} mixtures suggest that from a chemical point of view, 
core erosion could operate in Jupiter \citep{Soubiran2016, Soubiran2017}.
This also depends crucially on the state of the core. \citet{Mazevet2019b} show that during Jupiter's history, the constituents 
that make up Jupiter's heavy-element core were at least at some point molten, and as such could be either mixed into the
envelope or become stably stratified due to semi-convection (e.g. 
\citet{Wood2013}). It is yet to be investigated whether convective mixing can indeed lead to core erosion during 
Jupiter's evolution. Semi-convection (or double-diffusive convection) has been proposed to explain that
Jupiter could harbor a large composition gradient \citep{Leconte2012}. In this region, the thermally unstable region would
not develop large-scale convection due to a stabilising mean molecular weight gradient. In the double-diffusive 
regime, diffusion of temperature is more efficient than that of composition, and therefore the composition can be stably
stratified.

Recent work has focused on investigating whether it is possible that a composition gradient in Jupiter's envelope is
stable against large-scale convection throughout its evolution \citep{Vazan2018}. A particular model that has a 
large dilute core, and matches some observational constraints was presented. However, the initial model was created
ad-hoc and was not based on formation models. This \corr{raised}{raises} a challenging question: what are the possible formation and 
evolution paths that lead to a dilute core as inferred from structure models of Jupiter?

Here, we \corr{follow}{followed} the formation of Jupiter in the core-accretion framework \citep{Mizuno1980,Pollack1996}, starting at early 
times before Jupiter acquired its gaseous envelope. Then, we \corr{follow}{followed} the runaway gas accretion phase and the subsequent 
evolution, properly accounting for mixing of heavy elements within the planet. This allows us to determine whether a primordial composition 
gradient could be formed and sustained in Jupiter to explain its fuzzy core, or if other mechanisms must be invoked.
We find that forming a fuzzy core in Jupiter, as suggested by recent studies, is challenging in the standard framework for Jupiter's 
formation. This paper is structured as follows:
in \S \ref{sec:methods}, we give an overview of how we modelled the thermal evolution of Jupiter with MESA 
\citep{Paxton2011,Paxton2013,Paxton2015,Paxton2018,Paxton2019}, and present the different formation scenarios we considered.
In \S \ref{sec:results}, we present our formation and evolution models, and show that the outer convection zone can quickly grow and
mix large parts of the envelope. We show that our results are robust in terms of our model parameters, which include opacity, equation of  
state, and \corr{allowing for semi-convection}{whether we allow semi-convective mixing} in \S \ref{sec:sensitivity}. In \S 
\ref{sec:discussion}, we put our results in the context of current interior structure models, and we summarise our findings in \S 
\ref{sec:summary}.

\section{Methods}\label{sec:methods}
\subsection{The MESA code}\label{sec:mesa}
We modelled Jupiter's thermal evolution from late-stage formation until today with the Modules for Experiments in Stellar Astrophysics 
(MESA) code, release 10108 \citep{Paxton2011,Paxton2013,Paxton2015,Paxton2018,Paxton2019} with a modified equation of state following 
\citet{Vazan2013} (see \S \ref{sec:mesa_eos} and Appendix \ref{sec:equation_of_state} for details). The code numerically solves the fully 
coupled 1D structure and evolution equations under hydrostatic equilibrium \citep{Paxton2011}. The equations are solved with the Henyey 
method \citep{Henyey1965} on an adaptive Lagrangian grid with the mass coordinate $m$. Energy transport by radiative diffusion, electron 
conduction, and convection is included. 

The local stability of a radiative layer is calculated with the Ledoux criterion $\nabla_T < \nabla_{ad} + (\varphi / \delta) \nabla_{\mu,}$ 
where $\nabla_{T} = d\ln T / d\ln P$, $\nabla_{ad}$, and $\nabla_{\mu} = d\ln \mu / d\ln P$ are the temperature gradient, adiabatic 
temperature gradient, and mean molecular weight gradient, respectively, and $\varphi = (\partial \ln \rho / \partial \ln \mu)_{P,T}$ and 
$\delta = (\partial \ln \rho / \partial \ln T)_{P,\mu}$ are material derivatives. In radiative regions, $\nabla_{T}$ is equal to the 
radiative temperature gradient $\nabla_{rad}$. In convective regions, the temperature gradient is obtained by solving the cubic mixing-length theory equations 
(see e.g. \citet{Kippenhahn2012} for details). To avoid partial pressure derivatives that can
be numerically noisy, MESA calculates the composition term as a quantity measuring the pressure difference of neighbouring shells due to 
composition \citep{Paxton2013}. If $(\varphi / \delta) \, \nabla_{\mu} > 0$, the denser material is below the 
lighter one, and the composition term is stabilising against convection. In a homogeneous region, $\nabla_{\mu} = 0$ and the Ledoux 
criterion reduces to the standard Schwarzschild criterion. If a region is Ledoux stable but Schwarzschild unstable, a double-diffusive 
instability (semi-convection) may be present that could lead to additional mixing \citep{Wood2013,Radko2014}. 
Apart from in \S \ref{sec:sensitivity}, we do not include semi-convection \corr{modelling}{} in this work.

\corr{Convection is treated}{In 1D planetary evolution models, convection is commonly treated} as a diffusive process with the mixing length recipe, which requires a free parameter $\alpha_{mlt}$, which, following
\citet{Vazan2018}, is set to 0.1. For semi-convection, we \corr{use}{used} the diffusive 
approximation from \citet{Langer1983,Langer1985} as implemented in MESA. This recipe also requires a free parameter, which can be 
interpreted as the height of the double-diffusive layers. The parameter's value is poorly constrained, and could range over a few orders of 
magnitude \citep{Leconte2012}.

For the opacity, we used two different sets of tables: in the inner regions, where pressures and temperatures are high, MESA includes the 
conductive opacity from \citet{Cassisi2007}. 
Although these opacity tables extend into the range relevant to planetary interiors, they were originally developed for stellar interiors,
where matter is fully ionised. As a result, the \citet{Cassisi2007} values should be taken with caution (see e.g. \citet{Podolak2019}
for discussion). As a result, we investigated the sensitivity of our results on the assumed conductive opacity in \S 
\ref{sec:sensitivity}. For the gas opacity at lower temperatures, we used the tables from \citet{Freedman2008}. The Freedman opacity 
does not include the effect of grains. As a test, we calculated our models including grain opacity using the analytical fit to the 
\citet{Ferguson2005} dust opacity with an extrapolation for the lower temperature as presented
by \citet{Valencia2013}. While the detailed evolutionary path and the final interior structure are affected by the inclusion of grains or 
scaling of the conductive opacity, we find that our general conclusions are unchanged (see \S \ref{sec:sensitivity} for details).

\subsection{A planetary equation of state for MESA}\label{sec:mesa_eos}
While MESA includes the SCvH \citep{Saumon1995} tables, the equation of state (EoS) implemented in MESA is not applicable 
for modelling giant planets with heavy-element mass fractions $Z > 0$. We therefore implemented a modified version of 
the EoS presented in \citet{Vazan2013} that is suitable for planetary interiors into MESA.
Our EoS combines the EoS from \citet{Chabrier2019} for \corr{hydrogen/helium}{hydrogen and helium} and QEoS \citep{More1988} for \corr{water/rock}{water or rock} with the additive
volume law to calculate thermodynamic variables for an arbitrary mixture. Details of the implementation are given in Appendix 
\ref{sec:equation_of_state}. While there are more advanced of equations of state for water for planetary interior conditions 
\citep{French2009,Mazevet2019a}, they do not cover the temperature-pressure range required for modelling Jupiter's formation and evolution. 
In addition, the water EoS we are using \corr{has been tested}{was} and applied for planetary evolution in previous work 
\citep{Vazan2013,Vazan2015,Vazan2016,Vazan2018}. In order to make sure that our conclusions are unaffected by the choice of the EoS, we 
compared the density at a given \corr{pressure/temperature}{pressure and temperature} for water with QEoS and the EoS from \citet{French2009}. We \corr{found}{find} that the QEoS 
underestimates the density by $\sim 5 - 10 \%$ in the relevant regime. This difference is significantly smaller than the differences that arise 
when we assume a different chemical composition for the heavy elements. The sensitivity of our results to the assumed EoS is presented in \S 
\ref{sec:sensitivity}.

\subsection{Initial model and runaway gas accretion}\label{sec:initial_model}
We \corr{use}{used} a modified MESA routine (create\_initial\_model) to create an initial adiabatic model corresponding to Jupiter
at the onset of the runaway accretion phase. At that point, proto-Jupiter has a mass of $58 \, M_{\oplus}$ and an entropy
$S \simeq 8\ k_B$ per baryon, which is roughly within the constraints given by formation models of Jupiter \citep{Cumming2018}.
The value of the total mass \corr{is}{was} chosen such that the primordial composition gradient can be extended and isn't constrained to
just a few Earth masses. The initial entropy for proto-Jupiter at the onset of runaway gas accretion depends on the formation model 
parameters. We purposely chose a low-entropy case in order to build a steep entropy gradient as proto-Jupiter accretes gas, which 
should suppress convection, and can assist in forming composition gradients.

\begin{figure}
    \centering
    \resizebox{\hsize}{!}{\includegraphics{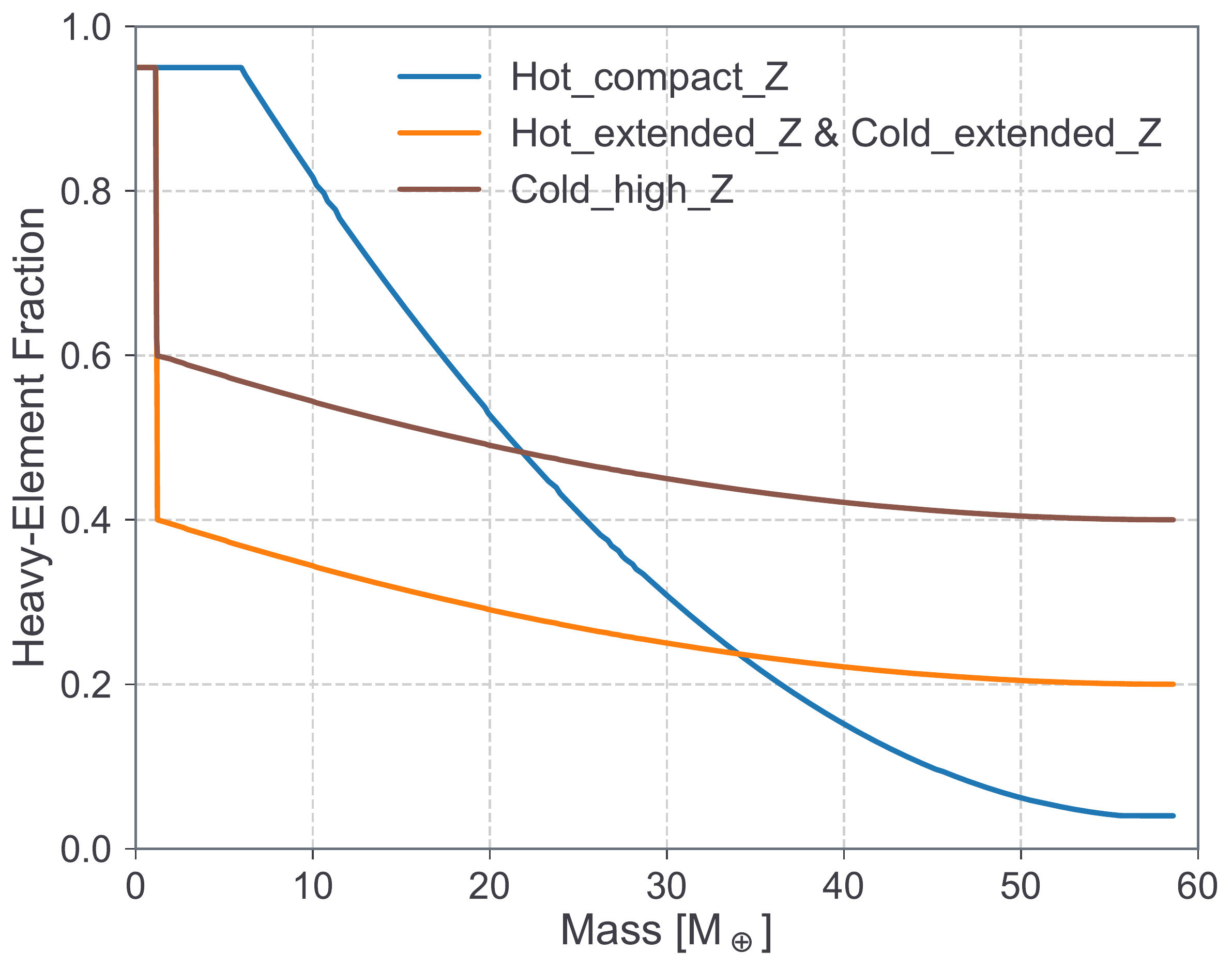}}
    \caption{Heavy-element distribution in proto-Jupiter vs.~mass at the onset of runaway gas accretion for the different models. The total mass in each model is $M = 58 \, M_{\oplus}$, and the total heavy element masses are $M_{Z} = 23$ (Hot\_compact\_Z) $16$ (Hot/Cold\_extended\_Z) and  $28 \, M_{\oplus}$ (Cold\_extended\_Z).}
    \label{fig:initial_composition}
\end{figure}

\begin{table*}
    \centering
    \begin{tabular}{cccccccc}
    \hline\hline
    & $M_Z$  [M$_{\oplus}$] & $S_{ph} \; [k_b / m_u]$ & $R$ $[R_J]$ & $ \dot{M}_{max} \; [M_{\oplus} / \textrm{yr}]$ & $\chi$ & $Z_{acc}$ & $L_{f} \; [L_{\odot}]$ \\
    \hline
    Hot\_compact\_Z & 23 & 8.1 & 1.2 & $10^{-2}$ & 1 & 0.04 & $5 \times 10^{-4}$ \\
    Hot\_extended\_Z & 16 & 6.5 & 1.3 &$10^{-2}$ & 1 & $\propto (1 - M(t) / M_{J})^{2}$ & $2 \times 10^{-4}$ \\
    Cold\_extended\_Z & 16 & 6.5 & 1.3 &$10^{-4}$ & 0 & $\propto (1 - M(t) / M_{J})^{2}$ & $1 \times 10^{-6}$ \\
    Cold\_high\_Z & 28 & 5.0 & 1.0 & $10^{-4}$ & 0 & $\propto (1 - M(t) / M_{J})^{2}$ & $1 \times 10^{-6}$ \\
    \hline
    \end{tabular}
    \caption{Summary of our models at the onset of the runaway accretion phase ($M_{tot} = 58 \, M_{\oplus}$) and the accretion parameters. The columns are the total heavy-element mass $M_{Z}$, photospheric entropy $S_{ph}$ per baryon, planetary radius in units of Jupiter's radius, max. accretion rate $\dot{M}_{max}$, shock temperature parameter $\chi$, the heavy-element fraction of the accreted material $Z_{acc,}$ and the post-formation luminosity $L_{f}$. We note that the difference in entropy between models is mainly due to the composition. The accretion parameters are chosen such that Hot-compact\_Z \& Hot\_extended\_Z create hot-start planets, while Cold\_extended\_Z Cold\_high\_Z create cold proto-Jupiters. The composition of the accreted material is either kept constant (Hot\_compact\_Z), or is changing as Jupiter grows. This creates a composition profile similar to that of \citet{Vazan2018} (Hot\_extended\_Z \& Cold\_extended\_Z) or with a much larger fraction of heavy elements in the envelope (Cold-high\_Z).}
    \label{tab:models}
\end{table*}

The composition in the interior \corr{is}{was} gradually adjusted in a pseudo-evolution (with the relax\_initial\_composition routine), such that 
the heavy-element fraction increases towards the centre. Since the true composition gradient of proto-Jupiter is unknown, we created 
initial models with three different composition gradients. Figure 1 shows the heavy-element distribution in 
proto-Jupiter  before the onset of runaway gas accretion. Hot\_compact\_Z roughly matches 
composition profiles from formation models of Jupiter that include planetesimal dissolution \citep{Lozovsky2017,Helled2017}, which 
generally leads to a steep gradient. Models Hot/Cold\_extended\_Z represent proto-Jupiters with a small dilute core that accretes a large 
amount of heavy elements during runaway gas accretion. The goal of these scenarios is to create a similar composition gradient to the 
starting model in \cite{Vazan2018}, who found that this initial condition can lead to a model of Jupiter that matches Jupiter's radius and
inferred moment of inertia today. In our work, we \corr{impose}{imposed} this composition profile by allowing for the accretion of heavy elements during 
runaway gas accretion. This \corr{is}{was} done by adjusting the composition of the material that is accreted as a function of the growing Jupiter's 
mass. In particular, we \corr{calculate}{calculated} the heavy-element mass fraction of the accreted gas as $Z_{acc} \propto (1 - M(t) / M_{J})^{2}$,
where $M(t)$ is proto-Jupiter's mass at time $t$, while keeping hydrogen-to-helium ratio constant at the proto-solar value,
meaning $X_{proto} = 0.705, Y_{proto} = 0.275$. The same is true for the third profile Cold\_high\_Z, except that it begins with more 
heavy elements and also accretes a much larger amount. We \corr{consider}{considered} this model in order to explore a formation pathway that leads to 
significant accretion of heavy elements during runaway. The heavy-element masses before runaway are $M_{Z}= 23, 16$ and $28 \, 
M_{\oplus}$, respectively. The heavy-element mass that is accreted during runaway is $\simeq 10 M_{\oplus}$ (Hot\_compact\_Z),
$\simeq 30 M_{\oplus}$ (Hot/Cold\_extended\_Z) and $\simeq 60 \, M_{\oplus}$ (Cold\_high\_Z), respectively.

Recent interior models  suggest that the maximum total heavy-element mass in Jupiter is $\simeq 45 M_\oplus$. Most of our models exceed this 
value (see \cref{tab:summary}). The reason is that the total heavy-element mass depends on the assumed chemical constituents for
the 'metals', and our calculations use a pure water EoS. It is therefore clear that our models require a higher heavy-element content.
In addition, our models are intentionally high in metals, since we focused on formation scenarios that
are favourable to yielding a dilute-core structure.

To model the growth of giant planets during the detached phase, we followed previous work by \citet{Berardo2017a}. During this stage, 
(for the most part) hydrogen and helium gas falls onto an accretion shock at the planet's surface \citep{Marleau2017,Marleau2019}.
The limiting gas accretion rate $\dot{M}_{g}$ was calculated using the formula from \citet{Lissauer2009} (their Eq. 2). 
Once the accretion rate reached a maximum value of either $10^{-2} \, M_{\oplus} / \textrm{yr}$ (Hot\_compact\_Z \& Hot\_extended\_Z)
or $10^{-4} \, M_{\oplus} /\textrm{yr}$ (Cold\_extended\_Z \& Cold\_high\_Z), we kept it constant at these values.
It is important to properly account for the accretion shock. Previous work \citep{Berardo2017a,Cumming2018} has shown that,
depending on the radiative efficiency of the shock, Jupiter \corr{could form}{could have formed} with an extended radiative envelope.
We adopted the shock temperature parameter $\chi$ of \cite{Cumming2018} that determines the temperature of the newly accreted material. 
It is zero if the planet's surface temperature is not affected by the accretion, and unity if the temperature of the 
accreted material \corr{radiates away the accretion energy}{reflects the fact that it is radiating away the accretion energy}.
In order to investigate the influence of a hot/cold runaway stage, we 
allowed Hot\_compact\_Z \& Hot\_extended\_Z to accrete at a maximum rate of $10^{-2} \, M_{\oplus} / \textrm{yr}$ with $\chi = 1$,
which represents a hot start. In Cold\_extended\_Z Cold\_high\_Z we set $\dot{M}_{max} = 10^{-4} \, M_{\oplus} / \textrm{yr}$ and
$\chi = 0$, yielding cold starts. The various models are summarised in Table \ref{tab:models}.
After reaching Jupiter's mass, runaway gas accretion was stopped, and we followed the long-term planetary evolution and 
the mixing of heavy elements in the interior.

\section{Results}\label{sec:results}
Firstly, we present the results of the four different formation models (see Tab. \ref{tab:models}). We \corr{calculate}{calculated} the 
evolution of these models for 4.5 Gyrs (Jupiter's age) and \corr{explore}{explored} whether Jupiter's interior becomes fully mixed or if
the dilute core or composition gradients can be sustained. Then, we present the final models at Jupiter's current age, and compare them with structure models of Jupiter.

\subsection{Formation models of Jupiter}\label{sec:format_models_of_jupiter}
The formation time from the onset of rapid gas accretion is $2.6 \times 10^{4}$ (Hot\_compact\_Z \& Hot\_extended\_Z)
and $2.6 \times 10^{6}$ (Cold\_extended\_Z \& Cold\_high\_Z) years, respectively.

\begin{figure}
    \centering
    \resizebox{\hsize}{!}{\includegraphics{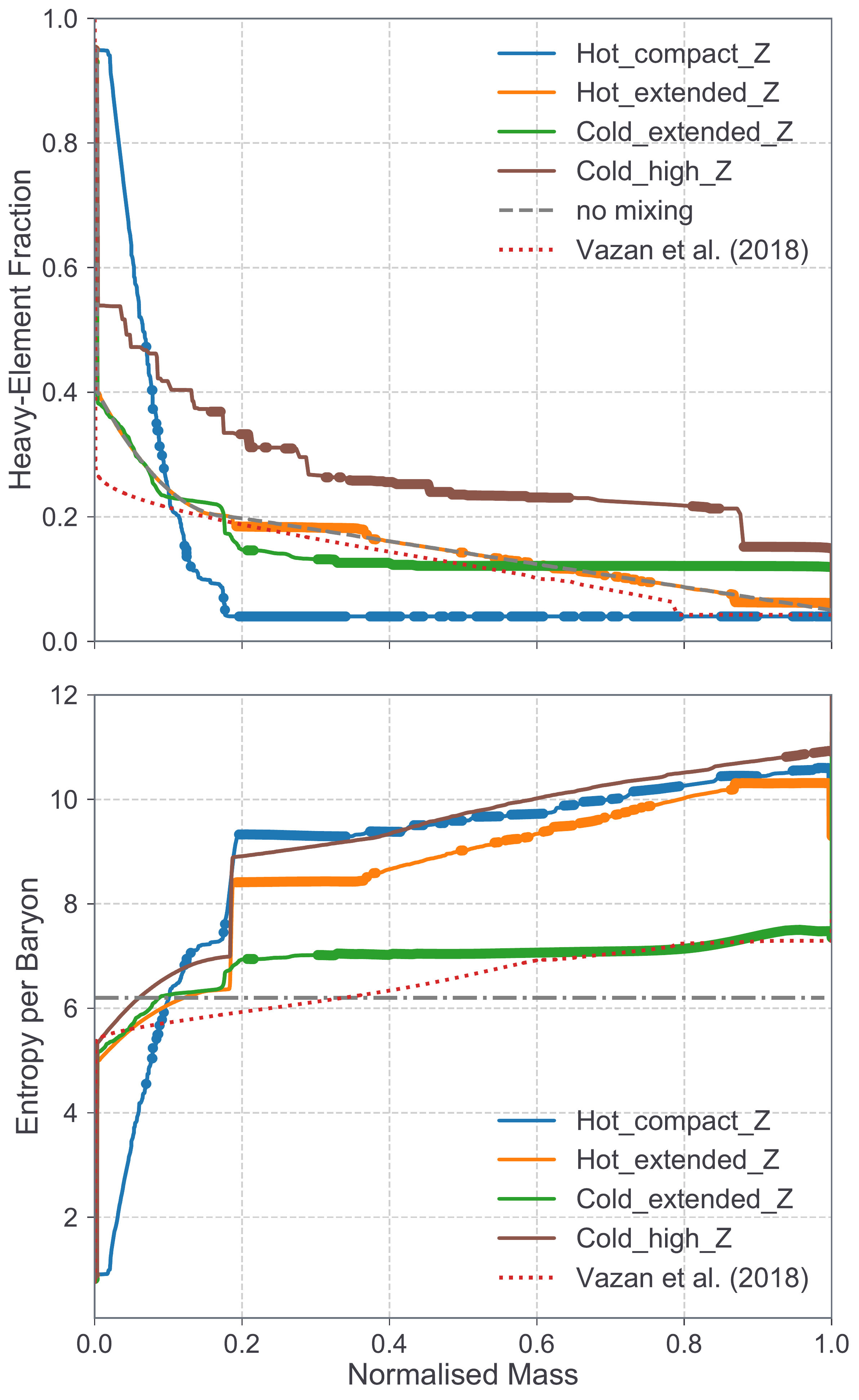}}
    \caption{Composition (top) and specific entropy (bottom) vs.~normalised mass at the time when proto-Jupiter reached its final mass (at $2.6 \times 10^{4}$ (Hot\_compact\_Z \& Hot\_extended\_Z) and $2.6 \times 10^{6}$ years (Cold\_extended\_Z \& Cold\_high\_Z). In both figures, thicker lines depict convective regions. For comparison, we also show how the accreted heavy-element profile without mixing (\corr{dashed grey}{grey dashed}) for Hot\_extended\_Z, as well as composition and entropy from \citet{Vazan2018} (\corr{dotted red}{red dotted}). The black dash-dotted line depicts roughly current Jupiter's entropy $S \simeq 6.2 \, k_{B}$/baryon \citep{Marley2007}.}
    \label{fig:accretion_composition_entropy}
\end{figure}

In \cref{fig:accretion_composition_entropy}, we present
proto-Jupiter's composition and entropy profiles at the time when runaway terminates. The deposition of the accretion
shock's energy in Hot\_compact\_Z \& Hot\_extended\_Z creates a large entropy gradient in the envelope, inhibiting convection
at early times. This is not the case in Cold\_extended\_Z \& Cold\_high\_Z, where the accretion rate is lower, and the
planet's surface temperature is unaffected by the shock. In Cold\_extended\_Z, this leads to vigorous convection 
and mixing in large parts of the envelope. Because the heavy-element mass accreted in Cold\_high\_Z is much higher,
the envelope is stable against large scale convection at this point. Similarly, there are no large convection zones in
Hot\_compact\_Z \& Hot\_extended\_Z. For comparison, we also show how the accreted composition profile would look if mixing
were shut off (same accretion as Hot\_extended\_Z, dashed line), yielding a very similar profile to the actual result of 
Hot\_extended\_Z. 

We also compare our results to the model presented in \citet{Vazan2018}. Our models have slightly steeper
heavy-element gradients. More importantly, they are much hotter by $\gtrsim 50$ \% throughout most of the envelope,
as demonstrated by the entropy profiles. Even our cold-start cases have a much higher entropy in the envelope, despite
the higher heavy-element content. For a rough comparison, we show the current entropy of Jupiter $\simeq 6.2\ k_B$ per baryon 
\citep{Marley2007}. This clearly shows that, in all of our models, there is energy available to mix the heavy elements into the envelope.

It should be noted that our models cover the expected range of luminosities calculated from  giant planet formation models
(see e.g. Figs 11 and 12 of \citet{Berardo2017b}). Even our coldest model yields a Jupiter with a higher entropy than inferred
from \citet{Vazan2018}. In our models, Jupiter forms rather hot with central temperatures of $\sim$ 50,000 K, where
in some cases the inner envelope is even hotter. This has clear consequences for understanding the long-term planetary
evolution and the expected luminosity of young giant planets.

\subsection{Jupiter's evolution after formation}\label{sec:jupiters_evolution}
\begin{figure*}
    \centering
    \includegraphics[width=17cm]{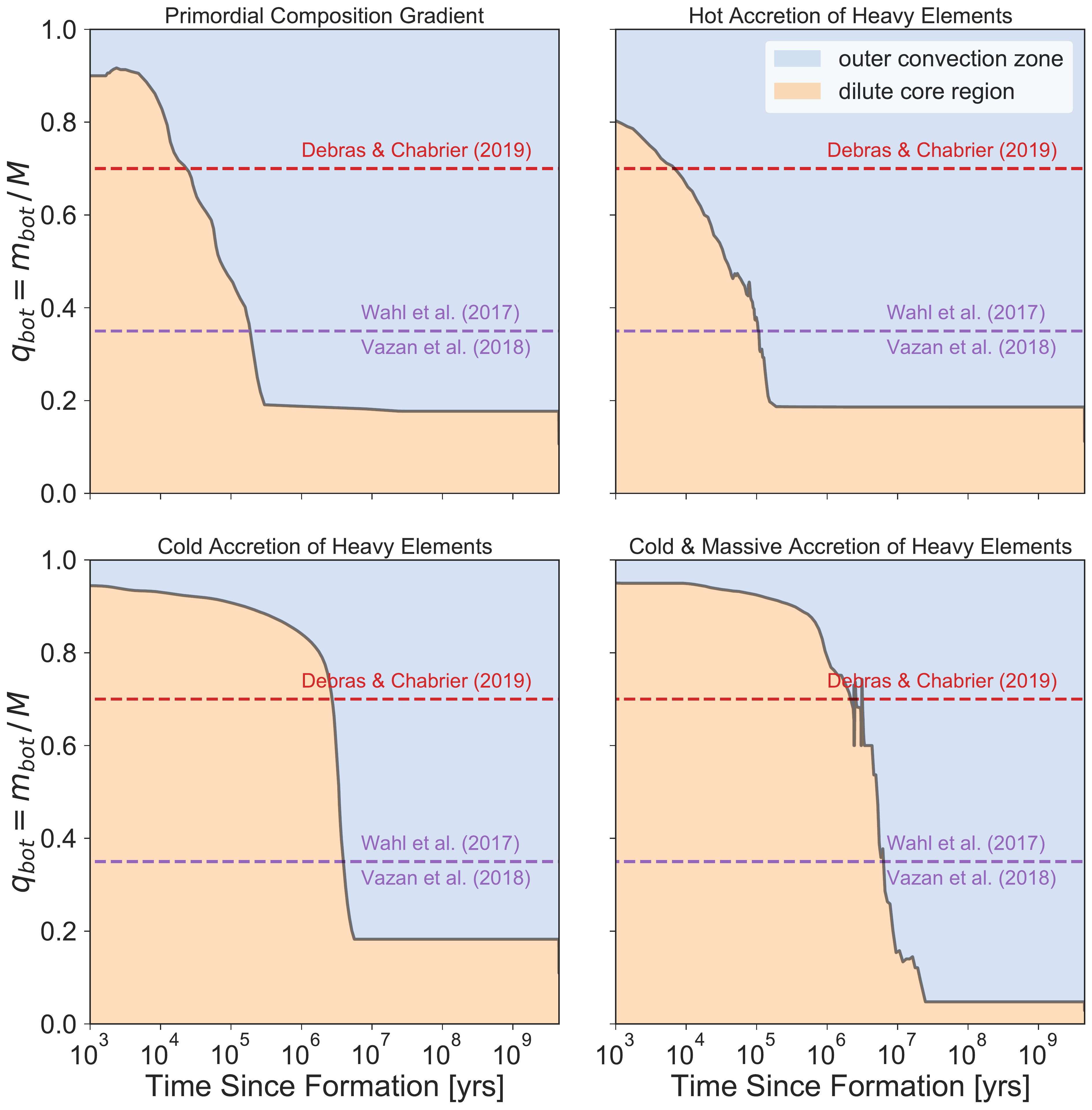}
    \caption{Mass coordinate at the bottom of the largest convection zone ($q_{bot} = m_{bot} / M$) as Jupiter evolves. We define the age to be zero at the point where Jupiter reaches its final mass and gas accretion has stopped. Curves are smoothed slightly in order to increase readability.}
    \label{fig:evolution_conv_mx1_bot}
\end{figure*}

Inhibiting convection at early times can be significant for Jupiter today. This is because mixing is most efficient 
when the planet is young and hot, and convective luminosities are large. In addition to the radiative envelope created
due to the accretion shock, a composition profile could further stabilise the envelope against large-scale convection.
In \cref{fig:evolution_conv_mx1_bot}, we show the location of the bottom of the largest convection zone
($q_{bot} = m_{bot} / M_J$) as Jupiter evolves and contracts. Below this mass coordinate there could still be small
convection zones, but no large-scale convection that thoroughly mixes the heavy elements into the envelope. We denote the 
region below the largest convection zone as the \textit{dilute core} region in \cref{fig:evolution_conv_mx1_bot}.
For comparison, the dashed lines depict the boundary of the fuzzy core as presented in structure models.

Hot\_compact\_Z \& Hot\_extended\_Z have similar behaviours. As Jupiter contracts and cools down, the convection zone quickly 
propagates inwards in $\sim 10^{5}$ years, until it stops when it reaches the steep, primordial composition gradient
($q_{bot} \simeq 0.2$). Despite the accreted heavy-element gradient in Hot\_extended\_Z, the convection zone advances at a
similar pace. This demonstrates that the heavy-element gradient is insufficient to stop large-scale convection in the envelope.
The reason is the large amount of thermal energy that is available to drive convection. Cold\_extended\_Z begins slightly more
convective (see \cref{fig:accretion_composition_entropy}). In the first few thousand years, the convection zone retreats,
because the mixing of heavy elements transiently stabilises the envelope. After that, the behaviour is very close to that of the
other models, except that it takes longer ($\sim 10^{6}$ years) to reach the stalling point. This is because there is less
thermal energy available since the accretion was slow and the surface temperature was not affected by the shock.
Cold\_high\_Z behaves very similarly, except that the outer convection zone is smaller at a given time compared to the
other models. Also, there can be large dredge-up events, where heavy elements from the deep envelope are mixed into the
upper envelope, shutting down convection for a short time.

All our models lead to a similar outcome: after, at most, a few tens of millions of years, the envelope 
is mostly mixed down to $q_{bot} \lesssim 0.2$, and there is only a small dilute-core region still intact.
This implies that standard formation models cannot lead to an extended fuzzy core.

The propagation and stalling of the convection zone can be understood in the following way, as discussed previously in \citet{Vazan2018}.
The convective part of the envelope is cooling faster than the radiative/conductive region, as energy is transported to the surface by
the convective flux. The boundary is characterised by an increasingly destabilising temperature gradient, and the convection zone moves 
inwards. 
 
However, at higher density, thermal conduction becomes more efficient, and the system can reach a quasi-steady state in which
the conductive heat transport across the boundary is sufficient to carry the full cooling luminosity. The difference in
temperature between radiative and convective shells stops growing, never becoming large enough to overcome the stabilising 
composition in the deep interior, and the innermost part of the dilute core is stable on a Gyr timescale. 

There are likely to be additional physical mechanisms not included in our calculations that act to mix heavy elements across
the thin boundary layers that separate convective regions. For example, turbulent motions could at least partially penetrate
the boundary and act to transport heavy elements across the interfaces \citep{Canuto1998,Canuto1999,Herwig2000,Moll2017}.
However, this would only further dilute the composition profile and allow the outer homogeneously mixed region to penetrate
even deeper into the interior. Including additional mixing mechanisms would therefore only strengthen our conclusions about
the difficulty of making an extended core.

\begin{figure}
    \centering
    \resizebox{\hsize}{!}{\includegraphics{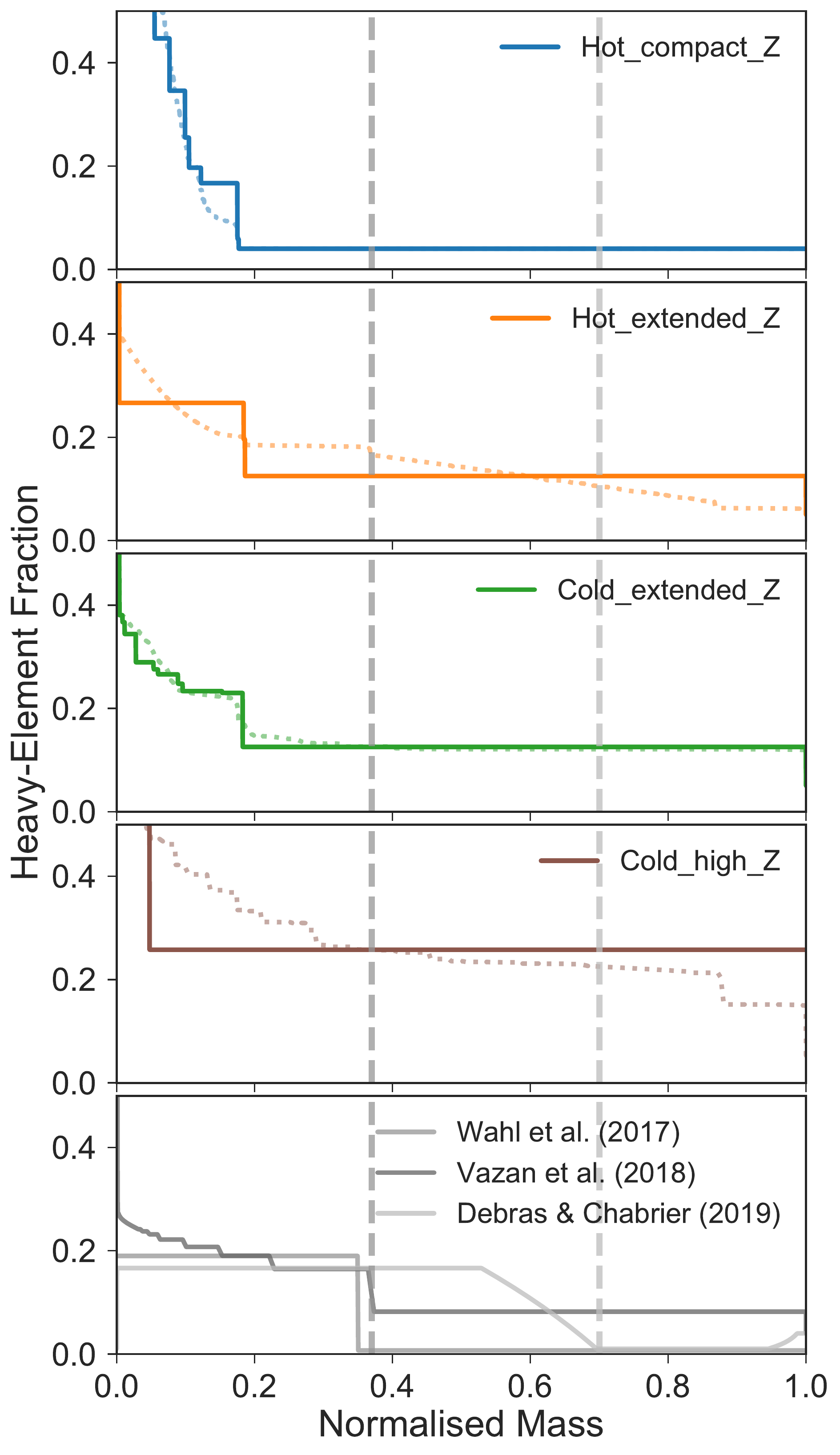}}
    \caption{
    Heavy-element mass fraction vs.~normalised mass of Jupiter shortly after its formation (dashed coloured lines) and today (solid coloured lines). The different panels (and colours) show the different formation pathways we consider. The grey vertical dashed lines indicate the extent of the dilute-core region as inferred by recent  evolution \citep{Vazan2018} and  structure models \citep{Wahl2017,Debras2019}. In all of our models, the dilute-core region is substantially smaller by $\lesssim 50\%$ by mass. The bottom panel shows  the heavy-element  distribution of these published models (solid grey lines).}
    \label{fig:o16_stacked}
\end{figure}

\subsection{Final structure}\label{sec:comparison_to_interior_models}
As shown in \cref{fig:evolution_conv_mx1_bot}, in all the cases we consider, the outer convection zone never reaches the deepest 
interior.
This is demonstrated in \cref{fig:o16_stacked}, which shows the heavy-element distribution as a function of normalised mass of 
our models at Jupiter's current age, compared to those of \citet{Wahl2017}, \citet{Vazan2018}, and \citet{Debras2019}.
The extent of the dilute core can be identified as the location of the outer-most significant jump in the metal content,
which in our models occurs at a mass of $m \lesssim 0.20 \, M_J$, which translates roughly into $r \lesssim 0.40 \, R_J$.
Hot\_compact\_Z shows a stable configuration where the primordial composition gradient was modified by the formation of
several small convection zones in the deep interior, resulting in a structure with many convective-conductive interfaces.
This can be identified by the several jumps in composition. The heavy elements from the primordial gradient are barely mixed into
the envelope, and the metallicity at the 1-bar level remains at $Z_{1bar} = 0.04$, as set by the accretion.

\begin{figure}
    \centering
    \resizebox{\hsize}{!}{\includegraphics{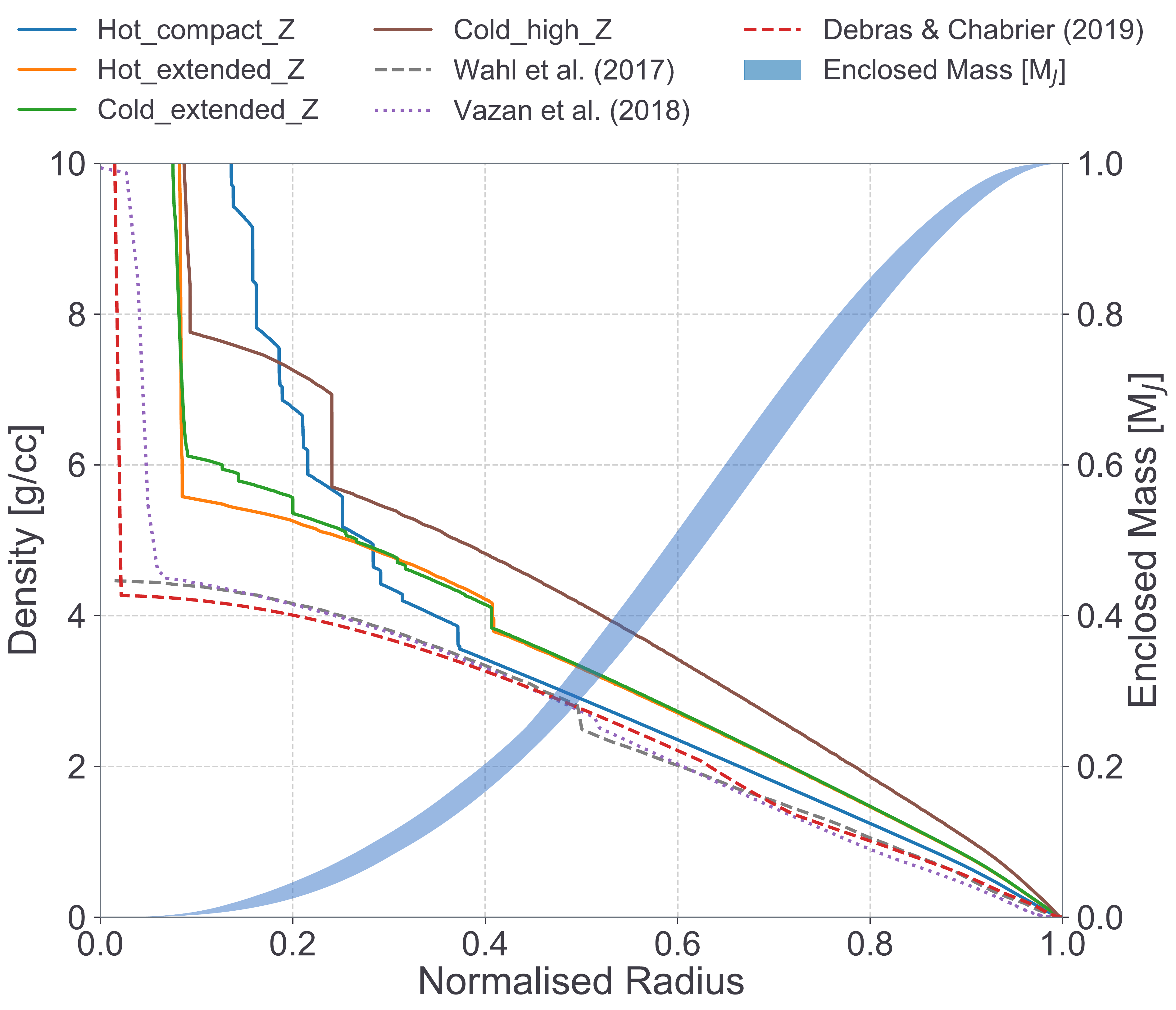}}
    \caption{Density vs.~normalised radius for our four Jupiter models (solid), and the dilute-core models of \citet{Wahl2017} (dashed grey), \citet{Vazan2018} (dotted purple) and \citet{Debras2019} (dashed red). The shaded blue region shows the range of possible enclosed masses for all density profiles.}
    \label{fig:density_radius_final}
\end{figure}

While the models Hot\_extended\_Z and Cold\_extended\_Z have very different accretion histories, the final structure is similar. 
The envelope is homogeneously mixed down to a radius of $r \simeq 0.4 \, R_J$, where a thin convective-conductive interface
separates two convection zones. Additionally, some of the heavy elements of the primordial composition are mixed into the outer 
envelope, increasing the metallicity of the outer envelope to $Z_{1bar} = 0.12$. It should be noted that the dilute core is 
roughly the same size in all models, including Cold\_high\_Z. This is because the mass of the steep primordial composition 
gradient is constrained by the planet's mass at the onset of rapid gas accretion. By comparison, in \citet{Wahl2017},
\citet{Vazan2018}, and \citet{Debras2019}, the dilute core occupies a region larger than $0.5 \, R_J$. 
The inferred density profiles from our evolution models are shown in 
\cref{fig:density_radius_final}. The density in our models is higher than the one inferred by structure models since 
they are colder, consist of more heavy elements, and do not satisfy all the observational properties of Jupiter today. 
It should be noted that these models are not meant to fit Jupiter's interior exactly. The goal of this study is to
explore formation paths that can lead to an extended dilute core. As a result, formation with high accretion rates of
heavy elements are preferred. 

Our models are summarised in Tab. \ref{tab:summary}, where we list the radius, total heavy-element mass, calculated normalised
moment of inertia (MoI), effective and 1-bar temperatures, and the heavy-element fraction at 1 bar. None of our models are
within the observational uncertainties of Jupiter's observed values. 
This is not surprising, since our Jupiter models are highly enriched in metals. Additionally, our model makes a number
of necessary simplifications. Therefore, even if we had the right amount of heavy elements, our models would not lead to Jupiter exactly. 
Among other things, the heavy elements are represented by pure water, the planet is assumed to be spherical and 
non-rotating, and our atmospheric model is simple. In addition, we do not consider helium rain \citep{Stevenson1975,Fortney2004,Wilson2010}.
However, the pressures and temperatures at the dilute-core boundary are far outside the H-He separation regime 
\citep{Morales2009,Morales2013,Schoettler2018}, and therefore helium rain is not expected to affect the characteristics of the
dilute core. Helium rain is more relevant for Saturn, since H-He separation occurs in the deep interior, closer to the planetary centre.

Another mechanism that could affect the heavy-element distribution within the planet is core erosion. The details of
this mechanism, however, are uncertain \citep{Guillot2004} and depend on the material properties
\citep{Wilson2012,Soubiran2016}. 
In the simple models of \citet{Guillot2004}, core erosion is seen as the entrainment of the core material into
the convection zone. A fraction of the kinetic energy in convection is used to mix heavy elements from the core
into the convective envelope.
If convective mixing is efficient, the material coming from the core
would likely homogeneously mix into the envelope instead of forming a dilute core. It is plausible that at later stages
in Jupiter's evolution, convection is less vigorous, and the eroded core material could create a composition gradient
where semi-convection operates. However, the details of this process are beyond the scope of this paper and require a 
detailed investigation. We hope to address this in future research.

\subsection{Sensitivity of the results to model assumptions}\label{sec:sensitivity}
In this section, we investigate the sensitivity of our results to the assumed model parameters including the EoSs, opacities, 
and the possibility of semi-convection. We take model Cold\_extended\_Z as the reference case. We show that while there are 
small changes in the heavy-element distribution, our conclusion that the extent of the dilute core region cannot extend beyond 
$\sim 20 \%$ of the planetary mass is robust.

\subsubsection{Equation of state}\label{sec:sensitivity_eos}
As discussed in \S \ref{sec:methods}, in all the models presented above, the heavy elements \corr{are}{were} represented by water. 
Denser materials (e.g. rocks, metals) have higher molecular weights, and therefore are harder to mix, and are more effective in 
suppressing convection \citep{Vazan2015}. Since the heavy elements in giant planets could include refractory materials, the 
models above might overestimate the efficiency of mixing. We therefore also ran models in which the heavy elements were 
represented by SiO$_2$ (rock). The SiO$_2$ EoS \corr{is}{was} again the QEoS from \citet{More1988}. 
In addition, we created a 50/50 mixture of rock \& water by combining their respective EoS, assuming ideal mixing. 
Since Jupiter's composition is dominated by 
hydrogen and helium, we also present a model in which the H-He EoS is calculated with the SCvH \citep{Saumon1995}
table instead of the newer version of \citet{Chabrier2019} (denoted by CMS) that we have implemented for the baseline cases.  
Figure 6 shows the heavy-element distribution in Jupiter today 
when using different EoSs, as described above. While the EoS clearly has an effect on the mixing in the deep interior, the 
location of the dilute core is unaffected.

\begin{figure}
    \centering
    \resizebox{\hsize}{!}{\includegraphics{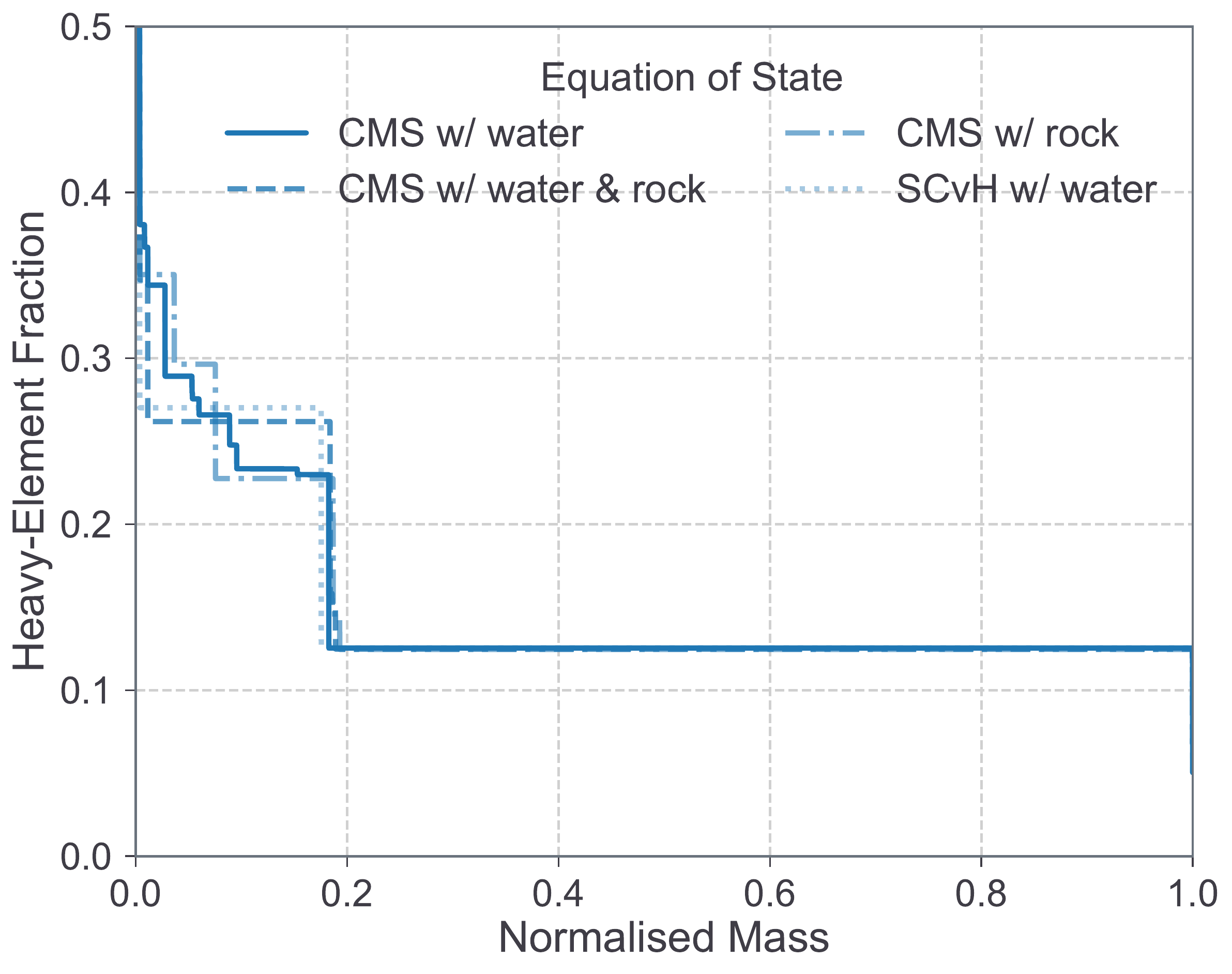}}
    \caption{Heavy-element mass fraction vs.~normalised mass in Jupiter today. The dashed line depicts our reference model Cold\_extended\_Z calculated with the CMS EoS for H/He \citep{Chabrier2019} and QEoS for water \citep{More1988}. The various solid lines are the same model calculated with different EoSs (see \S \ref{sec:sensitivity_eos}).}
    \label{fig:composition_sensitivity_eos}
\end{figure}

\subsubsection{Opacity}\label{sec:sensitivity_opacity}
\begin{figure}
    \centering
    \resizebox{\hsize}{!}{\includegraphics{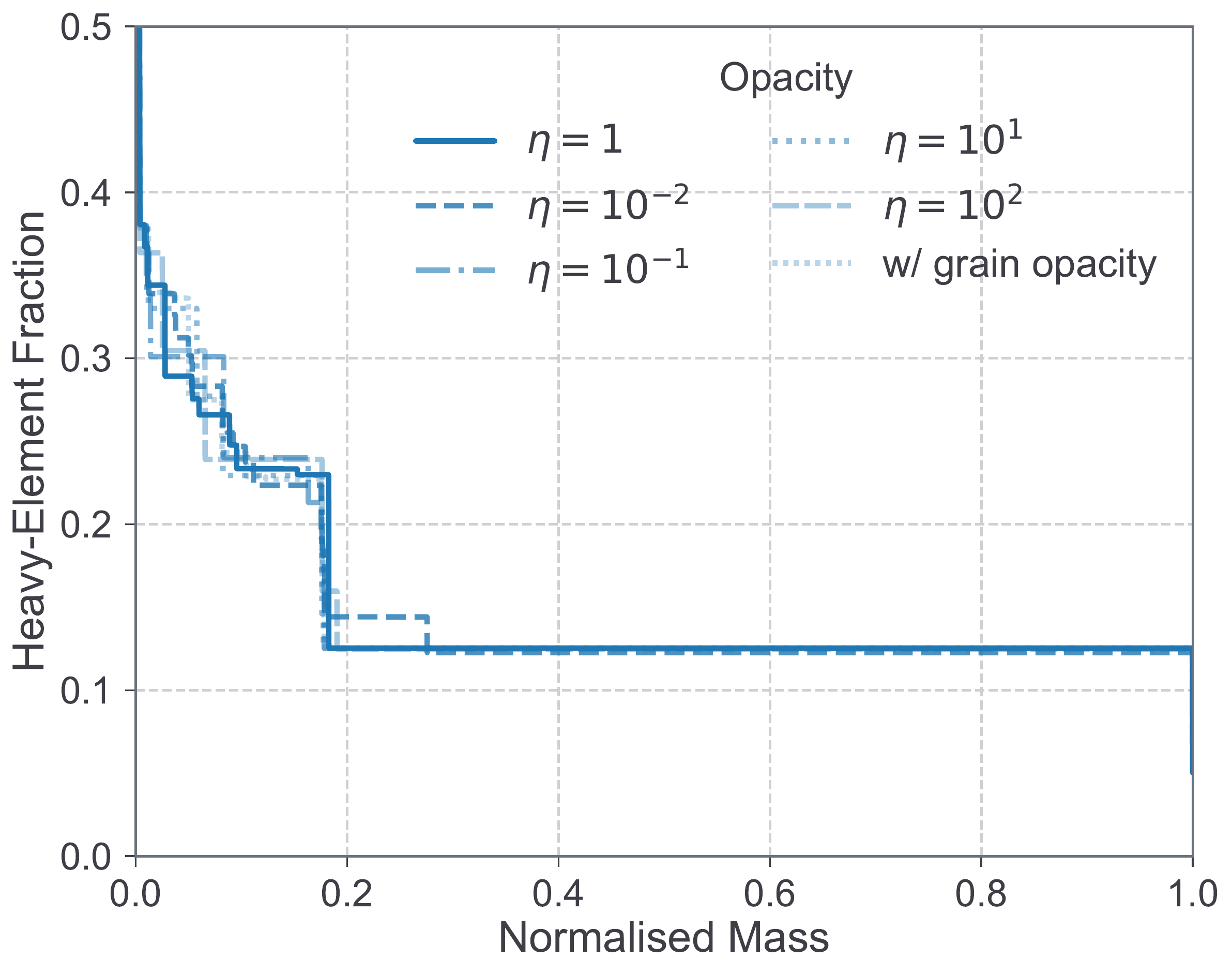}}
    \caption{Heavy-element fraction vs.~normalised mass in Jupiter today. The dashed line depicts our reference model Cold\_extended\_Z calculated without the grain contribution to the opacity. The various solid lines are the same model calculated with either the conductive opacity scaled by a factor $\eta$ or with grain opacity included (see \S \ref{sec:sensitivity_opacity}).}
    \label{fig:composition_sensitivity_opacity}
\end{figure}

Another important ingredient in the evolution calculation is the assumed opacity and whether it includes the
contribution from grains since this can have a strong influence on the planetary cooling. As a result, the 
mixing in the interior and the extent of the dilute core can also be affected. To test the robustness of our results
to the assumed opacity, we re-calculated the evolution of Cold\_extended\_Z once with grain opacity included (see \S 
\ref{sec:mesa}). We also separately scaled the conductive opacity by a factor $\eta = 0.01, 0.1, 10, 100$. In 
\cref{fig:composition_sensitivity_opacity}, we show the inferred  heavy-element profile in Jupiter today for the 
different models. Although the exact distribution of the heavy elements depends on the assumed opacity, the location of the 
dilute core remains unchanged. The only model where the location of the dilute core shifts is when the conductive opacity is 
reduced by a factor of 100. This is because an increase in the conductivity stabilises the conductive/convective interface, and 
the energy can be transported through the interface more efficiently.

We \corr{found}{find} that an increased opacity generally leads to the stairs being less stable. However, there is no qualitative 
difference in the final heavy-element profile when the conductive opacity is  increased or decreased by a factor of 10. As a result, 
even if the conductive opacity used in this work is not ideal for Jupiter's interior, it suggests that our main conclusion rather 
insensitive to the assumed opacity.

\subsubsection{Semi-convection}\label{sec:sensitivity_semiconvection}
\begin{figure}
    \centering
    \resizebox{\hsize}{!}{\includegraphics{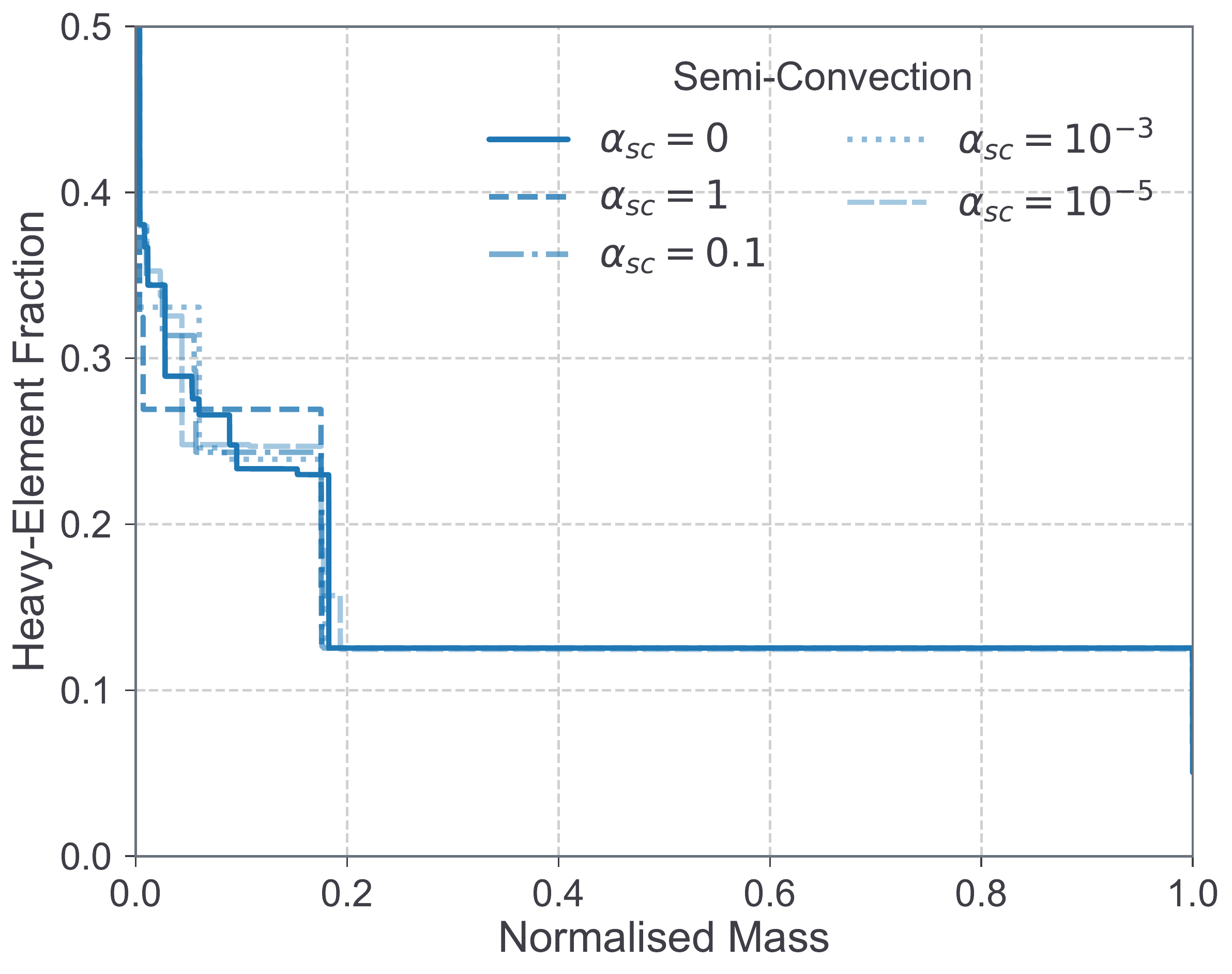}}
    \caption{Heavy-element fraction vs.~normalised mass in Jupiter today. The dashed line depicts our reference model Cold\_extended\_Z calculated without semi-convection. The various solid lines are the same model calculated with semi-convection included and a specific $\alpha_{sc}$ (see \S \ref{sec:sensitivity_semiconvection}).}
    \label{fig:composition_sensitivity_semi}
\end{figure}

Our baseline models were calculated without considering semi-convection in Ledoux-stable but
Schwarzschild-unstable regions. Semi-convection could destabilise composition gradients
by allowing for an efficient transport of heat across double-diffusive layers (see \S \ref{sec:mesa}).
We therefore re-calculated Cold\_extended\_Z when enabling semi-convection with a range of  $\alpha_{sc}$, which can be
interpreted as the typical layer height \corr{}{(measured in pressure scale heights)} of the double-diffusive regions. The resulting heavy-element profile is presented in 
\cref{fig:composition_sensitivity_semi}. We find that including semi-convection does not yield
significantly different heavy-element profiles in our models. In particular, the extent of the dilute core is largely 
unaffected.

\begin{table*}
    \centering
    \begin{tabular}{ccccccc}
    \hline\hline
     & Radius [R$_J$] & M$_Z$  [M$_{\oplus}$] & MoI & $T_{e}$ [K] & $T_{1bar}$ [K] & $Z_{env}$ \\
    \hline
     Jupiter & 1.0 & $\sim 10 - 65$ & 0.264 & 124 & 165 & $0.04$ \\
     Hot\_compact\_Z & 0.96 & 34 & 0.244 & 136 & 182 & 0.04 \\
     Hot\_extended\_Z & 0.92 & 49 & 0.235 & 117 & 153 & 0.12\\
     Cold\_extended\_Z & 0.92 & 49 & 0.237 & 116 & 153 & 0.12\\
     Cold\_high\_Z & 0.86 & 87 & 0.205 & 120 & 161 & 0.27\\
    \hline
    \end{tabular}
    \caption{Summary of our models at 4.5 Gyrs and measured/inferred values for Jupiter. The columns are the radius, total heavy-element mass, normalised moment of inertia (MoI), effective temperature $T_e$, 1-bar temperature $T_{1bar}$, and the heavy-element fraction in the homogeneously mixed region of the envelope $Z_{env}$. The basic observed properties and total heavy-element estimate are taken from \citep{Helled2017} and \citet{Wahl2017}. We note that in Hot/Cold\_extended\_Z and Cold\_high\_Z, the total heavy-element content is a combination of the primordial core and the heavy elements accreted during runaway, which are $\sim 33$ (Hot/Cold\_extended\_Z) and $\sim 60 \, M_{\oplus}$ (Cold\_high\_Z).}
    \label{tab:summary}
\end{table*}

\section{Connection to interior models of Jupiter}\label{sec:discussion}
Recent interior models of Jupiter suggest that the classic view of a simple core-envelope 
structure cannot be brought into accordance with observations \citep{Wahl2017,Debras2019}. Instead, in order to satisfy the  
gravity data, an internal structure with a dilute core seems to be required. In fact, there is no clear definition
of a \corr{'diluted/fuzzy core'}{'dilute or fuzzy core'}. What is usually meant is that there is a region in the innermost 
part of the planet that is enriched in heavy elements, possibly in the form of composition gradients, that extends 
far into the envelope \citep{Helled2017}.
The dilute core presented by \citet{Wahl2017} extends to $\simeq 0.5 \, R_J$, and is a region that is enriched with heavy 
elements by a constant factor compared to the outer envelope. In the models of \citet{Debras2019}, the dilute core corresponds to 
linearly decreasing the heavy-element fraction up to $\simeq 0.7 \, R_J$, going from $Z \simeq 0.2$ in the deep interior to a few 
times solar. Nevertheless, in both models, the dilute core spans a large 
fraction of Jupiter's radius (and mass). 

In both \citet{Wahl2017} and \citet{Debras2019}, the extent of the dilute-core region \corr{is}{was} 
adjusted so that the resulting density profile matches Jupiter's gravitational moments as measured by
Juno \citep{Folkner2017,Iess2018}. In \citet{Wahl2017}, the dilute-core boundary \corr{is}{was} modelled as a
step function-like drop in the heavy-element fraction, while \citet{Debras2019} used a linearly decreasing
function. In these models, the authors justify the stability of this region by invoking semi-convection
to this inner-most part. These two interior structure models also use a different EoSs for H-He, which leads to
a different envelope metallicity. An important point is that the location of the dilute core is empirical - it is
not a result of a physical process or the transport properties of the chemical elements.

Our study suggests that such an extended composition gradient is not a natural outcome of formation and
evolution models. We begin with a primordial composition gradient in the deep interior that was present
before Jupiter acquired its massive  gaseous envelope. As a result, the extent of the dilute core at
this point is limited by proto-Jupiter's mass at the onset of runaway gas accretion ($\simeq$ 20 \% of the mass in our models above). 
One may naively think that it should be possible to create a composition gradient that is less steep and spans
more of the planet's mass by, for example, slowly eroding the heavy-element gradient and constructing
a new, stable composition gradient that is less steep. However, our calculations show that this is not 
the case. As Jupiter's radiative envelope cools down and the convection zone pushes inwards, heavy elements
are transported outwards, but in that case they become homogeneously mixed across the entire outer convection zone. 
Alternatively, thermal conduction across the boundary can stabilise the boundary, as discussed in \S 
\ref{sec:jupiters_evolution}. In this case, however, we find that the interior composition gradient forms
a staircase of convective layers that maintain the overall mean composition gradient, so that the heavy elements 
remain in the innermost region.

Model Cold\_high\_Z was set to be very favourable to create a final structure with an extended dilute core.
Firstly, $\simeq 60 \, M_{\oplus}$ of heavy elements \corr{are}{were} accreted during runaway gas accretion to form a
massive composition gradient that spans the entire envelope by construction. Secondly, the accretion parameters
\corr{are}{were} chosen to yield a `cold' Jupiter to reduce the convective luminosity available for mixing. However,
even in this model, the outer regions quickly destabilise, since the stabilising effect of the composition
is insufficient to stop the outer convection zone from advancing inwards. Only the deep interior,
where the composition gradient is very steep, remains unmixed.

In \citet{Vazan2018}, the composition gradient remains mostly intact throughout Jupiter's lifetime,
and only the outer envelope becomes fully mixed. The difference between our results and the calculation
of \citet{Vazan2018} can be explained by the way in which the models are constructed. In \citet{Vazan2018},
the initial temperature and composition profile \corr{was}{were} chosen ad-hoc such that the final structure at 4.5 Gyrs matches 
observations. The formation process was not modelled, rather the evolution calculation began with a fully formed Jupiter.
As a result, their initial model is significantly colder than ours (see \cref{fig:accretion_composition_entropy}),
and therefore, mixing is less efficient. In this study, we followed Jupiter's formation and modelled the runaway phase,
which leads to higher entropy (and therefore hotter) envelopes. Even in our attempt to make the coldest possible models,
the entropy is still significantly higher than in \citet{Vazan2018}.

We therefore conclude that it is extremely challenging for standard formation models to arrive at a
dilute core solution as inferred from structure models that fit Juno's gravity measurements.
There are a few reasons for that. Firstly, the stable composition gradient is constrained to the deep interior
because it is already present at the onset of runaway gas accretion. 
Secondly, if the dilute core is a result of the runaway gas accretion, then it 
is not stable against convection unless proto-Jupiter is unrealistically cold, which seems rather unlikely 
\citep{Berardo2017a,Cumming2018}. It also requires extremely large amounts of heavy elements to be accreted,
which is not possible if Jupiter formed in-situ \citep{Shibata2020}. Thus, in order to reach a dilute core in Jupiter, additional mechanisms are required. One possibility
is a vast migration of Jupiter. Indeed, recent numerical simulations of planetesimal accretion that include
Jupiter's migration suggest that it could capture tens of $M_{\oplus}$ of heavy elements \citep{Shibata2020},
if the protoplanetary disc is sufficiently massive. This would mean that Jupiter must have migrated over a large
distance, for example starting at $\sim 10$ au. It should be noted, however, that this does not guarantee that 
the envelope would be stable against large scale convection, as demonstrated in case Cold\_high\_Z. 

Another potential path to form a dilute core is via a giant impact, as suggested by \citet{Liu2019}. In this scenario, 
a giant impactor ($\sim 10 \, M_{\oplus}$) provides the energy necessary to disrupt the primordial compact core and mix the 
heavy elements into the envelope. \citet{Liu2019} showed that under certain conditions (impact parameters and initial 
thermal state of Jupiter), the extended dilute core is stable over Jupiter's lifetime. In fact, our models suggest that such 
a giant impact must occur at a relatively later stage when Jupiter is cold enough to avoid significant mixing of the interior. 

Our results suggest standard formation paths are unlikely to create an extended dilute-core structure unless additional physical
processes such as layered convection \citep{Leconte2012} or core erosion \citep{Guillot2004,Moll2017}, a giant impact, or 
significant migration are included. Even for these scenarios, a detailed investigation is still needed, and a process that leads 
to the formation  of Jupiter's dilute core is yet to be presented.

\section{Summary}\label{sec:summary}
We calculated the formation and long-term evolution of Jupiter starting from the onset of runaway gas accretion until today, properly accounting for the
energy transport and heavy-element mixing in the deep interior. We investigated different formation scenarios, with primordial 
composition gradients and various heavy-element accretion rates and shock properties during runaway gas accretion. Our main 
findings can be summarised as follows:

\begin{itemize}
    \item If Jupiter accretes a homogeneous envelope on top of a primordial composition gradient of heavy-elements (model Hot\_compact\_Z), mixing in the deep interior leaves the original structure mostly intact. Almost none of the heavy elements are mixed into the envelope - the dilute core is too compact.
    \item If the composition gradient is the result of a large amount of heavy-element accretion during runaway, the envelope quickly ($\sim 10^{6}$ years) becomes unstable in the face of large-scale convection (models Hot\_extended\_Z \& Cold\_extended\_Z). The result is a fully mixed envelope, with a small dilute core in the deep interior.
    \item Even a cold and slow runaway gas accretion ($\tau_{acc} \sim 10^{6}$ yrs) combined with a very high accreted mass of solids ($\simeq 60 \, M_{\oplus}$) leads to a formation path where the outer convection zone destroys the accreted composition gradient (model Cold\_high\_Z).
    \item None of our models lead to a structure of Jupiter today with an extended dilute core. Current interior models require a dilute core that has an extent of $\simeq 0.5$--$0.7 \, R_{J}$, significantly larger than found in this study.
\end{itemize}

We conclude that forming a fuzzy core in Jupiter as suggested by recent studies is in fact very challenging in the standard 
picture of giant planet formation. One solution for Jupiter's extended dilute core is a giant impact. However, this must have 
occurred late, otherwise Jupiter would have been too hot, and convection would have been too efficient for the composition gradient to survive. 
Future works should focus on the correct modelling of additional processes that could influence the mixing, such as helium rain, 
double-diffusive convection, and core erosion, as well as the development of more sophisticated formation models.

\begin{acknowledgements}
We thank S.~Mazevet for the careful reading of the paper and valuable comments. We also thank D.~Stevenson, P.~Bodenheimer and A.~Vazan for fruitful discussions, and F.~Debras and G.~Chabrier for sharing their data with us. We also acknowledge support from SNSF grant \texttt{\detokenize{200020_188460}} and the National Centre for Competence in Research ‘PlanetS’ supported by SNSF.  AC is supported by an NSERC Discovery grant and is a member of the Centre de Recherche en Astrophysique du Qu\'ebec (CRAQ) and the Institut de recherche sur les exoplan\`etes (iREx).
\end{acknowledgements}

\bibliographystyle{aa}
\bibliography{library}

\appendix
\section{Equation of State}\label{sec:equation_of_state}
For the temperature/density regime relevant for planetary modelling, MESA includes the SCvH \citep{Saumon1995}
EoS tables for mixtures of hydrogen and helium. In this regime, however, the tables do not include
the contribution from heavy elements. A common approach to bypass this issue is to model the planet as consisting of
an inert core that is treated as an inner boundary condition for the structure equations, and a gaseous 
\corr{H/He}{hydrogen-helium} envelope (see, e.g. \citet{Malsky2020}). 
However, this approach does not work if one wants to follow the mixing of heavy elements, for
example. Therefore, we implemented an EoS into MESA that is 
suitable for planetary interior conditions and includes either water or rock as a heavy element.
Our EoS is an updated version of the EoS from \citet{Vazan2013}, where the two equations of state
SCvH \citep{Saumon1995} for hydrogen and helium and QEoS \citep{More1988} for water or rock (SiO$_2$)
were combined to allow for an arbitrary mixture of the three components, by replacing the SCvH with
the recent H-He EoS of \citet{Chabrier2019}.

The calculation of mixtures makes use of the additive volume law $1 / \rho(p, T, \vec{X}) = X / \rho_{H}(p, T) + Y / 
\rho_{He}(p, T) + Z / \rho_{Z}(p, T)$, where we denote the dependence on an arbitrary 
composition by $\vec{X}$ and X, Y, Z are the mass fractions of hydrogen, helium, and the heavy elements, respectively. 
The required thermodynamic quantities such as the entropy follow similar additive laws. We neglect the entropy of mixing,
as it is a small correction that does not have a significant impact on planetary evolution \citep{Baraffe2008}.

To implement this EoS into MESA, we created a grid of tables with compositions varying by $\Delta X, \Delta Z = 0.1$ 
in the range of $0 \leq Z \leq 1$ and $0 \leq X \leq 1$. The tables \corr{are}{were} calculated on a grid of $\log Q$ and $\log T$,
where $\log Q = \log \rho - 2 \log T + 12$. These tables are then read and used for the evolution calculations by the 
EoS module in MESA, with a few modifications. The module then calculates the required thermodynamic quantities for a
mixture of hydrogen, helium, and the heavy elements by interpolating between the tables. Water (and rock) undergoes phase 
transitions at certain relevant pressures and temperatures. This introduces discontinuities in the derivatives of 
thermodynamic variables. Since every entry in the tables needs to be valid, we \corr{smooth}{smoothed} over phase transitions 
with a cubic spline interpolation that uses the closest valid neighbouring points.

\begin{figure}
    \resizebox{\hsize}{!}{\includegraphics{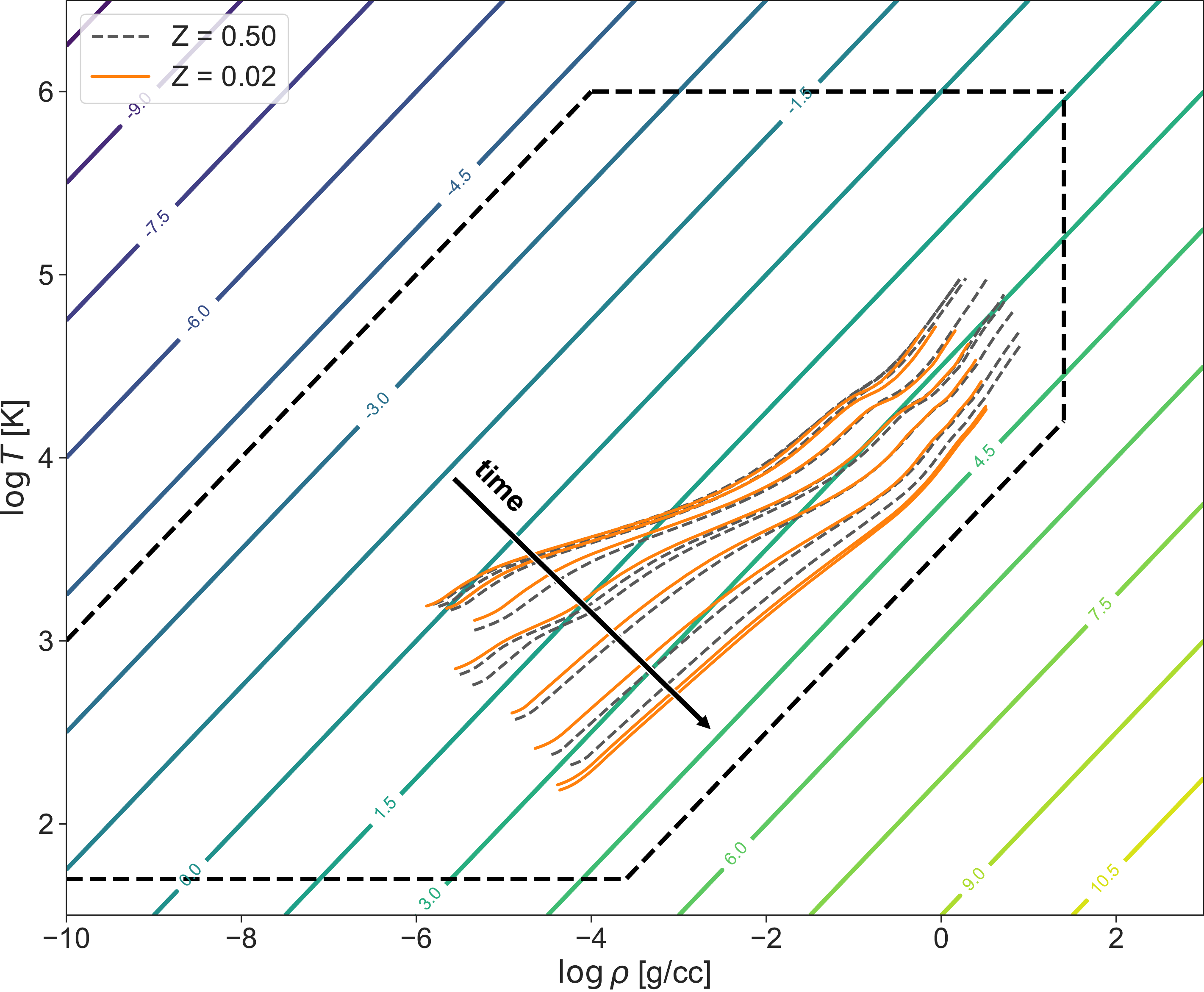}}
    \caption{Cooling of two homogeneous one Jupiter mass planets in the $\log \rho$-$\log T$ plane calculated with our version of MESA that includes the modified equation of state. One of them is metal-poor (orange), and the other is metal-rich (dashed grey). The contours correspond to values of $\log Q = \log \rho - 2 \log T + 12$. Dashed black lines show the boundaries beyond which the equation of state is not valid. The arrow indicates the direction of time.}
    \label{fig:eos_test}
\end{figure}

In \cref{fig:eos_test}, we show the cooling of two \corr{homogeneous planets and one Jupiter-mass planet}{homogeneously mixed one Jupiter-mass planets} calculated with MESA 
using our equation of state. The difference between the two models is only the bulk metallicity, which is $Z = 0.02,
0.50$, respectively. As expected, the metal-rich planet is much denser and hotter than the metal-poor one. Both 
cooling tracks are well within the equation of state boundaries (dashed black lines), demonstrating that the current 
range is sufficient.

As an additional test to see whether our EoS is working properly, we reproduced the results from 
\citep{Vazan2018} by calculating Jupiter's evolution starting with the same temperature and composition profiles. 
While there were some small differences in the final interior structure at Jupiter's age (due to different opacities,
resolutions, and numerical methods), we \corr{get}{got} a very similar result, in particular regarding the extent of the dilute core.
We therefore conclude that our EoS implementation in MESA is working reliably.

A limitation of our EoS implementation is that we do not currently include the effect of heavy elements in the compressional
heating term $\epsilon_g$ in the energy equation (see discussion in \S 2.5 of \citealt{Mankovich2016}). 
We find that there are numerical difficulties, because the extra derivative would have to be calculated by a simple two point estimation, which is 
numerically noisy and leads to unreliable results. We leave this for future works.

\end{document}